\documentclass{article}
\PassOptionsToPackage{authoryear}{natbib}
\usepackage[final]{neurips_2024}
\usepackage[utf8]{inputenc} %
\usepackage[T1]{fontenc}    %
\usepackage[hidelinks]{hyperref}       %
\usepackage{url}            %
\usepackage{booktabs}       %
\usepackage{amsfonts}       %
\usepackage{nicefrac}       %
\usepackage{microtype}      %
\usepackage{xcolor}         %
\usepackage{graphicx}
\usepackage{longtable}
\usepackage{caption}
\usepackage{mdframed}
\usepackage{subcaption}
\usepackage{multirow}
\usepackage{placeins}
\usepackage{multicol}
\usepackage{makecell}
\usepackage[normalem]{ulem} %
\usepackage{wrapfig}
\usepackage[percent]{overpic}
\usepackage{lipsum}
\usepackage{csquotes}
\usepackage[OT2,T1]{fontenc}
\usepackage[english]{babel}
\usepackage{devanagari}
\usepackage{tablefootnote}
\usepackage{pdfpages}
\usepackage{caption}
\usepackage{amsmath}
\usepackage{enumitem}

\usepackage{macros}

\usepackage{hyperref}  
\hypersetup{
     colorlinks=true,
     linkcolor=blue,
     citecolor=blue,
     filecolor=blue,
     urlcolor=blue,
 }

\title{Seed-TTS: A Family of High-Quality\\Versatile Speech Generation Models}

\author{Seed Team, ByteDance\thanks{Please cite this work as ``Seed-TTS (2024)''. The full statement of author contributions and acknowledgments can be found at the end of the document. Correspondence regarding this technical report should be sent to \href{mailto:Seed-TTS@bytedance.com}{Seed-TTS@bytedance.com}.}}

\begin{document}

\maketitle
\begin{abstract}
We introduce Seed-TTS, a family of large-scale autoregressive text-to-speech (TTS) models capable of generating speech that is virtually indistinguishable from human speech. Seed-TTS serves as a foundation model for speech generation and excels in speech in-context learning, achieving performance in speaker similarity and naturalness that matches ground truth human speech in both objective and subjective evaluations. With fine-tuning, we achieve even higher subjective scores across these metrics. Seed-TTS offers superior controllability over various speech attributes such as emotion and is capable of generating highly expressive and diverse speech for speakers in the wild.
Furthermore, we propose a self-distillation method for speech factorization, as well as a reinforcement learning approach to enhance model robustness, speaker similarity, and controllability. We additionally present a non-autoregressive (NAR) variant of the Seed-TTS model, named $\text{Seed-TTS}_\text{DiT}$, which utilizes a fully diffusion-based architecture. Unlike previous NAR-based TTS systems, $\text{Seed-TTS}_\text{DiT}$ does not depend on pre-estimated phoneme durations and performs speech generation through end-to-end processing. We demonstrate that this variant achieves comparable performance to the language model-based variant and showcase its effectiveness in speech editing. We encourage readers to listen to demos at \url{https://bytedancespeech.github.io/seedtts_tech_report}.
\end{abstract}

\section{Introduction}
\label{sec:intro}
We present Seed-TTS, a family of speech generation models capable of synthesizing speech with human-level naturalness and expressiveness. It can also create controllable, high-fidelity synthesized speech based on a short enrollment speech clip in a zero-shot manner. This model has significant potential in applications such as virtual assistants, audio books, video dubbing, and more. 

The primary goal of Seed-TTS is to create a speech generation model that approaches human-level speech, even for arbitrary speakers in the wild with little data. Seed-TTS has been evaluated on three tasks: zero-shot speech in-context learning (ICL), speaker fine-tuning, and emotion control. We release the configuration of our test dataset for future benchmarking and also discuss the model's behavior regarding product deployment.

We further introduce two novel extension techniques that can significantly enhance model performance: speech factorization via self-distillation and preference biasing through reinforcement learning (RL). For the former, unlike commonly applied methods such as feature engineering \citep{chen2023streaming,wang2024streamvoice,wang2023lm} or specialized loss formulations \citep{ju2024naturalspeech,lajszczak2024base} or model designs \citep{qian2019autovc,jiang2023mega}, our simple self-distillation scheme enables Seed-TTS to achieve high-quality timbre disentanglement without altering model structure or loss function. For the latter, we employ RL techniques \citep{kaelbling1996reinforcement,li2017deep} and demonstrate their effectiveness in improving robustness, speaker similarity, and controllability.


We then compare the advantages and disadvantages of two major categories for speech generation: language model-based \citep{wang2023neural,zhang2023speak,lajszczak2024base} and diffusion-based \citep{ju2024naturalspeech,gao2023e3,chen2022resgrad,lovelace2023simple} modeling. To this end, we designed a non-autoregressive (NAR) variant of Seed-TTS, named $\text{Seed-TTS}_\text{DiT}$, which is a fully diffusion-based speech generation model that directly predicts output speech latent representations in an end-to-end manner, rather than relying on a separate duration prediction module, as in previous NAR methods \citep{tan2022naturalspeech,le2024voicebox,du2024unicats,jiang2023mega,ren2019fastspeech,yi2022softspeech,yi2022prosodyspeech}. We show that $\text{Seed-TTS}_\text{DiT}$ performs comparably to autoregressive language model-based methods and demonstrate its speech editing capabilities.

Lastly, we discuss potential applications and limitations of Seed-TTS, as well as several challenges we encountered during development, including those related to building socially responsible artificial intelligence (AI). The capabilities and limitations of Seed-TTS give rise to significant and novel challenges in multimedia and safety applications that we believe must be carefully studied when considering their potential societal impact.

Our key contributions are as follows:
\begin{itemize}[leftmargin=2em]
    \item We introduce Seed-TTS, a family of speech generation models capable of generating highly expressive, human-like speech. We demonstrate that Seed-TTS achieves state-of-the-art (SOTA) performance in multiple evaluations. Under a zero-shot ICL setup, we show that Seed-TTS is able to generate robust, similar, and highly dynamic speech that is indistinguishable from human speech.
    \item We present a novel self-distillation extension of Seed-TTS for timbre disentanglement and demonstrate SOTA performance in the voice conversion task.
    \item We introduce a novel RL-based post-training extension for Seed-TTS, which holistically improves the model's performance.  
    \item We present a novel fully diffusion-based variant of Seed-TTS, which achieves superior generation quality. We show its advantages in the speech editing task and compare it with its language model-based counterpart.
\end{itemize}

\section{Method}
\label{sec:method}

\begin{figure}[h]
\centering
\includegraphics[width=0.9\textwidth]{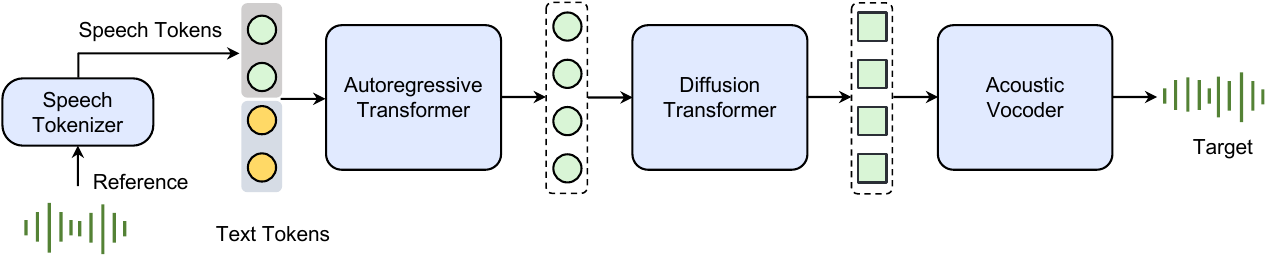}
\caption{\small An overview of the Seed-TTS inference pipeline. (1) The speech tokenizer learns tokens from reference speech. (2) The autoregressive language model generates the speech tokens based on the condition text and speech. (3) The diffusion transformer model generates continuous speech representations given generated speech tokens in a coarse-to-fine manner. (4) The acoustic vocoder yields higher-quality speech from the diffusion output.}
\label{fig:system}
\end{figure}

Seed-TTS is an autoregressive transformer-based \citep{touvron2023llama,vaswani2017attention} model, as depicted in \autoref{fig:system}. Our system consists of four main building blocks: a speech tokenizer, a token language model, a token diffusion model, and an acoustic vocoder. We emphasize that Seed-TTS is trained on large amounts of data (orders of magnitudes larger than the previously largest TTS systems) to enable strong generalization and emergent abilities.

First, a speech tokenizer converts the speech signal into a sequence of speech tokens, upon which a token language model is trained using a method similar to those described in \citet{betker2023better,lajszczak2024base}, and \citet{wang2023neural}. 
We investigate both continuous and discrete speech tokenizers, and found that the design of the tokenizer is crucial to the performance of the entire system.
The language model is trained on paired sequences of text and speech tokens. During inference, it generates speech tokens autoregressively. Note that in this technical report we focus on the speech generation task, so the loss for the text sequence is masked. These generated tokens are then processed by the diffusion model to enhance acoustic details. The output is passed to the acoustic vocoder to predict the final waveform. The acoustic vocoder is separately trained with a design similar to \citet{kumar2024high,lee2022bigvgan,cong2021glow} and \citet{liu2021basis}.

Similar to text-based language models, Seed-TTS undergoes three training stages: pre-training, fine-tuning, and post-training. The pre-training stage aims to maximize scenario and speaker coverage while establishing a robust backbone for general speech modeling. As mentioned before, Seed-TTS utilizes a volume of training data and model scale that are orders of magnitude larger than previous speech generation models during this stage.

The fine-tuning stage consists of speaker fine-tuning and instruction fine-tuning. Speaker fine-tuning focuses on enhancing performance for a selected group of speakers, whereas instruction fine-tuning aims to improve controllability and interactivity. Post-training is conducted through RL, which holistically improves the model.


We observe two major advantages of the Seed-TTS model compared to prior models. Firstly, Seed-TTS demonstrates superior naturalness and expressiveness in its speech synthesis capabilities across various scenarios, including challenging ones such as shouting, crying, or highly emotional speech. During development, we rigorously tested the model in scenarios that are considered difficult or impossible for previous TTS systems, showing clear advantages over prior SOTA systems. Examples are showcased in \S\ref{subsec:clone}.

Secondly, Seed-TTS addresses stability issues prevalent in language model-based TTS systems, which hinder their real-world deployment. Stability is achieved through a combination of token and model design improvements, enhanced training and inference strategies, data augmentation, and reinforcement post-training. Consequently, Seed-TTS achieves significantly better robustness across test sets. 


Serving as a foundation model for speech generation, Seed-TTS can perform various tasks, such as speech ICL, controllable TTS, cross-lingual TTS, voice conversion, timbre generation, and speaking style transfer. In this report, we demonstrate Seed-TTS in the tasks of speech ICL, speaker fine-tuning, controllable TTS, and voice conversion.

Specifically, our ICL results, also known as zero-shot voice continuation, are detailed in \S\ref{subsec:clone}. ICL is defined as generating a novel spoken utterance with the same timbre and prosody as a short reference speech clip \citep{wang2018style,wang2023neural,zalan2022audiolm}. The ICL results are obtained by continuing audio and text prompts with the pre-trained Seed-TTS model. Results of speaker fine-tuning and instruction fine-tuning are presented in \S\ref{sec:sft}, with reinforcement post-training results discussed in \S\ref{sec:rl}. Voice conversion results are presented in \S\ref{sec:vc}.

\section{Experiments}
\label{sec:exp}

\subsection{Zero-shot in-context learning}
\label{subsec:clone}
We prepare two test sets, denoted as \emph{objective-set} and \emph{subjective-set}, for these experiments. The objective set consists samples extracted from English (EN) and Mandarin (ZH) public corpora that are used to measure the model's performance on various objective metrics. Specifically, we employ 1,000 samples from the Common Voice dataset \citep{ardila2019common} and 2,000 samples from the DiDiSpeech dataset \citep{guo2021didispeech}. The subjective set consists of 100 samples in both English and Mandarin sampled from an in-house dataset used for subjective evaluation, containing significantly richer speech than the objective set, including highly expressive speech with diverse accents, dialects, emotions, and speaking styles.

For both test sets, we ensure that each sample contains one reference utterance and one target utterance spoken by the same speaker. The proposed Seed-TTS system is applied to generate speech of the target text based on the reference speech as an audio prompt. In this way, we can directly compare synthesized speech against ground truth speech from real humans. The duration of the reference utterance ranges from 3 to 20 seconds. 

\paragraph{Evaluation metrics.} We adopt the word error rate (WER) and speaker similarity (SIM) metrics for objective evaluation. For WER, we employ Whisper-large-v3 \citep{radford2023robust} and Paraformer-zh \citep{gao2023funasr} as the automatic speech recognition (ASR) engines for English and Mandarin, respectively. For SIM, we use WavLM-large fine-tuned on the speaker verification task \citep{chen2022large,Chen_2022_wavlm} to obtain speaker embeddings used to calculate the cosine similarity of speech samples of each test utterance against reference clips. We use Comparative Mean Opinion Scores (CMOS) studies for subjective evaluation, as follows. For each test sample, human evaluators are first shown a reference speech clip of the target speaker. They are then presented with the synthesized output of our model and the corresponding ground truth human speech, played in random order. Evaluators are asked to rate the sample with higher speaker similarity and expressiveness to the reference clip on a scale between -2 to +2, where -2 and +2 indicate the least and strongest preference for the first sample. We collect the results, rearrange each comparison in the order of ``Seed-TTS vs. Human'', and average the preference scores over all evaluators and test sentences. Empirically, an absolute CMOS score less than 0.1 is considered to be insignificant between two systems. The results for both test sets are reported in \autoref{tab:clone}. We release the configuration of the objective set in this \href{https://github.com/BytedanceSpeech/seed-tts-eval/tree/main}{GitHub repository} to enable benchmarking.\footnote{\small{Due to copyright restrictions, we are not releasing the subjective set. All samples in the demo page are included with authorization.}}


\begin{table}[h]
\centering
\begin{tabular}{l|c|c|c|c}
\hline
 \multirow{2}{*}{\textbf{System}}& \multirow{2}{*}{\textbf{Lang.}} & \multicolumn{2}{c|}{\textbf{Objective set}} & \textbf{Subjective set} \\ \cline{3-5}
 &  & \textbf{WER ($\downarrow$)} & \textbf{SIM ($\uparrow$)} & \textbf{CMOS ($\uparrow$) vs. Human} \\\hline
Seed-TTS & EN & 2.249 &\textbf{0.762} & -0.07 \\
Vocoder resynthesized & EN & 2.165 & 0.702 & - \\
Human & EN & \textbf{2.143} & 0.730 & - \\
\hline
Seed-TTS &ZH &\textbf{1.115}&\textbf{0.796}& -0.08\\
Vocoder resynthesized & ZH & 1.342&0.733&-\\
Human & ZH& 1.254& 0.750&-\\
\hline
\end{tabular}
\caption{\label{tab:clone}\small Evaluation results of Seed-TTS against resynthesized and real human speech.}
\end{table}

\paragraph{In-context learning results.} From \autoref{tab:clone}, we observe that Seed-TTS achieves a WER similar to ground truth human speech with significantly higher speaker similarity. This result may be explained by the observation that the ground truth and reference utterances can still differ in speaking style and background environment, even when spoken by the same speaker. In contrast, Seed-TTS accurately captures the characteristics of the reference speech when generating the target utterance, resulting in a more consistent and faithful reproduction of the enrollment clip. We showcase the ICL examples at \href{https://bytedancespeech.github.io/seedtts_tech_report/\#zero-shot-icl-samples}{this page}. 

It is noteworthy that a lower WER does not necessarily lead to an improved subjective score on speaker similarity. We empirically observe that a lower WER typically indicates that the model produces more ``standardized'' speech that is easier for the ASR system to recognize, but at the expense of other desirable qualities. For example, in cases where the prompt speech contains a strong accent or high expressiveness, obtaining a lower WER from generated speech usually indicates less accented speech with limited variation in the output space of the model, which may sound less natural and have reduced speaker similarity when measured in subjective evaluations.

In subjective tests, Seed-TTS achieves performance closely matching real human speech for both English and Mandarin with CMOS scores of -0.07 and -0.08, respectively. 
Note that the subjective test set includes diverse and expressive speech. During early development, we conducted the same evaluation on several prior models, such as \citet{jiang2023mega,le2024voicebox,wang2023neural,zhang2023speak,song2024ella,ren2020fastspeech,ju2024naturalspeech}, and \citet{shen2023naturalspeech}, all of which produced CMOS results below -1, indicating a substantial gap between synthesized and real human speech. The subjective test for Seed-TTS marks the first instance of a TTS system generating results indistinguishable to real human speech in a zero-shot ICL setting with in-the-wild speech prompts. 
For samples with lower CMOS scores, evaluators noted that real human speech contained more variations across sentences, while the synthesized speech maintained consistent prosody defined by the reference. This consistency leads to better similarity with the speech prompt but results in slightly fewer prosodic variations for long-form speech generation. The multi-shot ICL approach may address this limitation, which we will investigate in future work.

\paragraph{Comparison to traditional speaker fine-tuned TTS models.} We compare our zero-shot ICL system against a set of traditional FastSpeech-based \citep{ren2020fastspeech,liu2022controllable} speaker fine-tuned TTS models. We collected speech from 10 speakers, categorized into two groups: a ``common'' speaker set (7 speakers) consisting of average, everyday speech, and a ``hard'' speaker set (3 speakers) consisting of speakers with strong accents or very unique, exaggerated speaking styles, e.g., an electronic high-pitched chipmunk virtual character. For Seed-TTS, a randomly selected sentence with an average duration of 15 seconds was used as the voice prompt for each speaker. The full training set of each speaker (roughly 5 hours each) was used to fine-tune separate, well-trained, traditional TTS systems with a similar setup to the one described in \citet{liu2022controllable}.

\begin{figure}[h]
\centering
\includegraphics[width=0.8\textwidth]{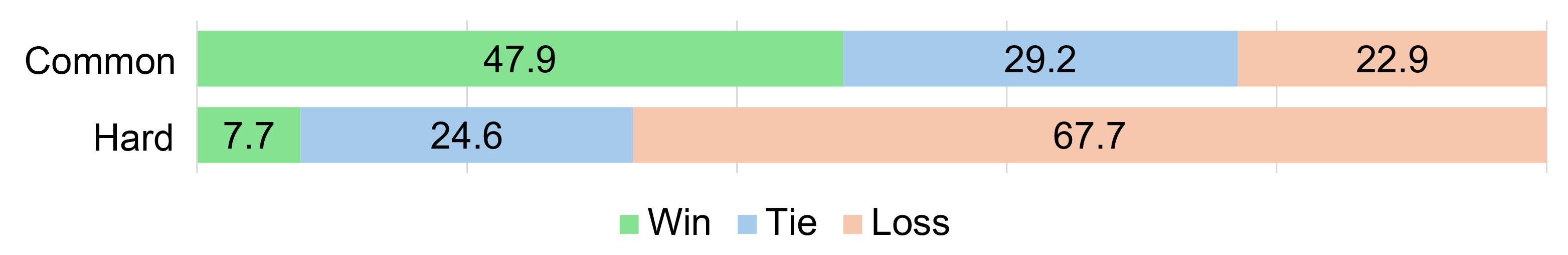}
\caption{\label{fig:bvs} \small Subjective preference between Seed-TTS zero-shot ICL (using 15s audio prompt) and traditional speaker fine-tuned neural TTS models (using 5 hours of data) using ``common'' and ``hard'' test sets.}
\end{figure}

For each speaker, 30 utterances were generated by each system, covering diverse scenarios, contexts, and emotions.
We measure the average preference rate of each system per speaker averaged from 10 human evaluators and present the results in \autoref{fig:bvs}.

We observe that for the ``common'' speaker set, our zero-shot ICL system was favored for 47.9\% of test samples over the traditional fine-tuned TTS systems. According to human evaluators, Seed-TTS demonstrated a clear advantage in naturalness and expressiveness. However, for ``hard'' speakers, the traditional fine-tuned model exhibited stronger performance. We speculate this is because the accents and unique speaking styles are not preserved as faithfully by our zero-shot ICL generation, particularly in cases where the representative prosody of the speaker was not included in the 15-second prompt. We believe with longer prompts and a better coverage of training data, such limitations can be alleviated. 


\paragraph{Speech understanding evaluation.} We further verify the generation quality of Seed-TTS by training an ASR model on generated speech \citep{le2024voicebox}. To this end, we generated a synthetic version of the LibriSpeech 960-hour training set \citep{panayotov2015librispeech} through a ``text-wave shuffling'' strategy and used the synthetic corpus to train an ASR model from scratch, which we then used to transcribe speech on the original LibriSpeech development and test sets. 
Specifically, we generate a synthetic version of each utterance in the training set by employing it as the audio prompt to synthesize a new sentence using randomly sampled text from the training set, while ensuring that all the utterances and text are sampled only once. In this way, we created a synthetic LibriSpeech training corpus which maintained the same total speaker and content information as the original corpus to train ASR models using the WeNet toolkit \citep{zhang2022wenet}. We adopted a 12-layer Squeezeformer \citep{kim2022squeezeformer} as the ASR encoder and a 3-layer bi-directional transformer as the ASR decoder. An ASR baseline model was also trained on the original LibriSpeech training corpus. All the models were trained using the same hyperparameters, e.g., number of epochs, batch size, learning rate, and so on. Each model was tested on the LibriSpeech development and test sets, the results of which are shown in \autoref{tab:tts-ls-asr}.

\begin{table}[h]
\centering
\begin{tabular}{l|c|c|c|c}
\hline
\textbf{Training data} & \textbf{\emph{dev\_clean}} & \textbf{\emph{dev\_other}} & \textbf{\emph{test\_clean}} & \textbf{\emph{test\_other}} \\ 
\hline
Synthetic data & 2.59 & 7.78 & 2.76 & 7.58 \\
Real data & \textbf{2.26} & \textbf{5.97} & \textbf{2.45} & \textbf{5.98} \\
\hline
\end{tabular}
\caption{\label{tab:tts-ls-asr} Comparison of WER ($\downarrow$) between models trained on synthesized data and real data in the ASR task.}
\end{table}

We observe that for the clean sets, i.e., \emph{dev\_clean} and \emph{test\_clean}, the model trained with synthetic data achieves very similar ASR performance to the model trained with real data. 1.81\% and 1.6\% absolute WER drops are observed on the noisy \emph{dev\_other} and \emph{test\_other} sets, respectively, which we speculate are due to Seed-TTS's tendency to reduce the background noise during the generation process, resulting in less robustness to noise. With data enhancement \citep{Chen_2022_wavlm,li2018developing}, we believe the gap will be reduced. This result suggests the potential for using synthetic data in the development of speech understanding models, which further pushes the unification of speech understanding and generation.

\begin{figure}[h]
\centering
\includegraphics[width=0.7\textwidth]{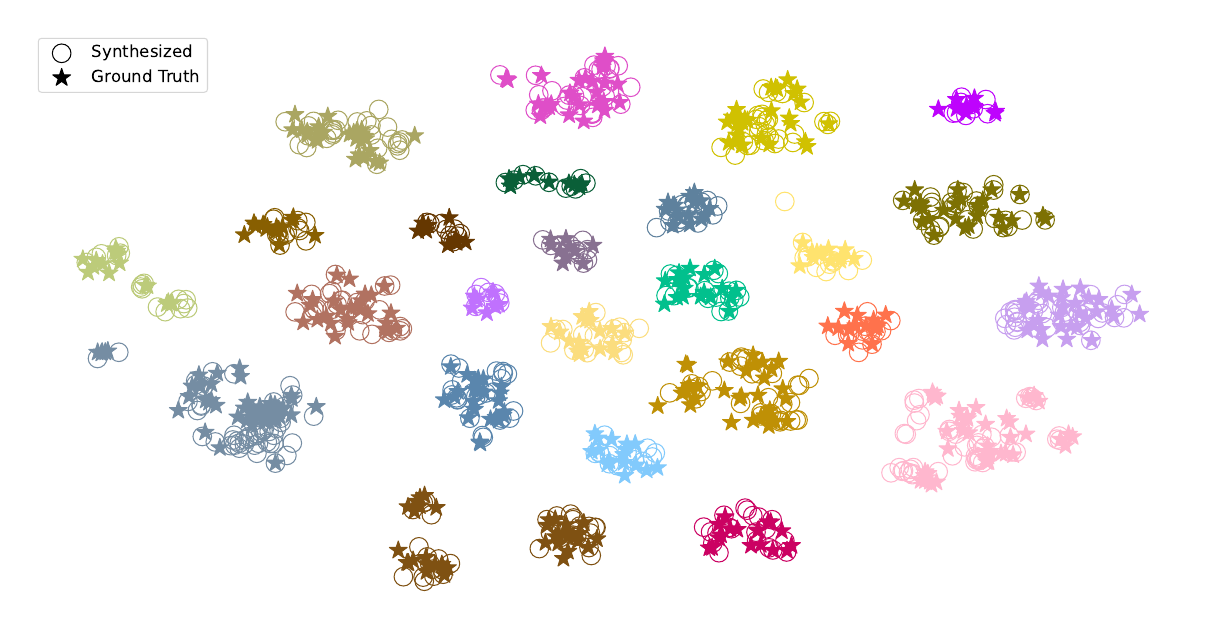}
\caption{\label{fig:vox}\small t-SNE visualization of speaker embeddings from the VoxCeleb1 test set (25 speakers) on synthesized and ground truth speech.}
\end{figure}

\paragraph{Visualizing speaker similarity of ground truth and ICL speech.} To verify the preservation of timbre in synthesized speech, we generated the English utterances from VoxCeleb1 test set \citep{nagrani2017voxceleb} using the same shuffling method as above and obtained their speaker embeddings using the WavLM-based speaker verification model from \citet{Chen_2022_wavlm}. We plot the speaker embeddings of ground truth and synthesized speech of 25 speakers using t-SNE \citep{van2008visualizing} in \autoref{fig:vox}.


We observe that the embeddings of ground truth and synthesized speech from the same speaker reliably cluster together, which supports the finding that the quality and speaker similarity of speech generated by Seed-TTS closely resembles real human speech.

\subsection{Speaker fine-tuning}
\label{sec:sft}
We perform speaker fine-tuning (SFT) on top of the base Seed-TTS pre-trained model. In this experiment, we selected 5 speakers (3 female and 2 male), each with speech data ranging from 1 to 10 hours. We fine-tuned Seed-TTS using their combined data, totaling 20 hours, and integrated an additional speaker index token to select the timbre of target speakers during inference. For these selected speakers, we evaluate the generated speech of the fine-tuned model ($\text{Seed-TTS}_\text{SFT}$) against that of the base pre-trained model ($\text{Seed-TTS}_\text{ICL}$) using WER and SIM objective metrics and subjective CMOS studies. For the base model, a randomly sampled voice clip of 20 seconds was used as the audio prompt for each speaker. The results of the speaker fine-tuning experiment are reported in \autoref{tab:comaprison_sft_icl}.


Compared to $\text{Seed-TTS}_\text{ICL}$, the fine-tuned model shows similar performance in objective metrics, but demonstrates an advantage in subjective evaluation with a CMOS score of +0.37. Our empirical observations indicate that the fine-tuned $\text{Seed-TTS}_\text{SFT}$ model captures more nuances of the target speaker, such as subtle prosody changes and distinctive pronunciation patterns at the end of sentences.

\begin{table}[h]
\centering
\begin{tabular}{l|c|c|c}
\hline
\textbf{System} & \textbf{WER ($\downarrow$)} & \textbf{SIM ($\uparrow$)} & \textbf{CMOS ($\uparrow$)} \\\hline
$\text{Seed-TTS}_\text{ICL}$ (Zero-shot in-context learning) & 3.15 & 0.779 & - \\
$\text{Seed-TTS}_\text{SFT}$ (Speaker fine-tuned) & \textbf{2.83} & \textbf{0.789} & \textbf{+0.37}\\
\hline
\end{tabular}
\caption{\label{tab:comaprison_sft_icl}Comparison between $\text{Seed-TTS}_\text{ICL}$ and $\text{Seed-TTS}_\text{SFT}$.}
\end{table}

\paragraph{Controllability through instruction fine-tuning.} To enable further controllability of our speaker fine-tuned model, we experiment with integrating additional instruction fine-tuning (IFT) \citep{yi2022prosodyspeech, yi2019singing, zhuang2021litesing, deng2023prosody}. IFT enables the model to flexibly control each aspect of generated speech such as expressiveness, speaking rate, style, emotion, and so on.
We showcase emotion control just as an example in this report.

To verify emotion controllability, we trained a speech emotion recognition (SER) model similar to \citet{Chen_2022_wavlm}, selected four primary emotions (i.e., angry, happy, sad, and surprised), and measured the accuracy of predicted emotions from synthesized speech. We generated and evaluated 100 utterances for each emotion, where the subject matter of synthesized text was designed to match the target emotion. 

The results are summarized in \autoref{tab:sft_emo_acc}. We find that even without an explicit controlling signal, $\text{Seed-TTS}_\text{SFT}$ still obtained moderate accuracy in emotion control. We speculate this is because the model has the capability to infer the appropriate target emotion based on the provided textual content. When combined with additional controlling signals, a significantly improved accuracy is obtained. The examples are demonstrated at \href{https://bytedancespeech.github.io/seedtts_tech_report/\#speaker-finetune-samples}{this page}.

\begin{table}[h]
\centering
\begin{tabular}{l|c|c|c|c}
\hline
\textbf{System} &  \textbf{Angry} & \textbf{Happy} & \textbf{Sad} & \textbf{Surprise} \\ 
\hline
 Seed-TTS$_\text{SFT}$  & 0.69 & 0.4 & 0.37 & 0.22 \\
 Seed-TTS$_\text{IFT}$ & \textbf{1.0}& \textbf{0.85} & \textbf{1.0} & \textbf{0.98} \\
\hline
\end{tabular}
\caption{\label{tab:sft_emo_acc} Comparison of emotion control accuracy ($\uparrow$) between $\text{Seed-TTS}_\text{SFT}$ and $\text{Seed-TTS}_\text{IFT}$.}
\end{table}



\subsection{Low-latency inference and streaming processing}
\label{sec:deploy}
The deployment of TTS models in real-world applications poses several practical challenges from multiple perspectives. For example, in chat-based applications the latency and first packet delay are essential for user experience. The computation cost in both time and memory are crucial for the serving concurrency. Compared with traditional TTS models, Seed-TTS adopts a significantly larger model size, creating additional barriers for deployment. To resolve these challenges, we employed various techniques to reduce inference cost and latency \citep{dao2022flashattention,ainslie2023gqa,luo2023latent,lin2023awq}. 
Specifically, we addressed three aspects for model deployment. Firstly, a causal diffusion architecture is implemented, which enables streaming processing in the diffusion module and significantly reduces the processing latency and first packet delay. Secondly, we employ consistency distillation \citep{song2023consistency} and a modified flow matching algorithm \cite{esser2024scaling} to reduce the computation cost of the diffusion model. On the other hand, we investigate commonly applied methods to reduce the memory and computation consumption on the language model side, such as grouped-query attention \citep{ainslie2023gqa}, paged attention \citep{kwon2023efficient}, flash attention \citep{dao2022flashattention,dao2023flashattention}, and model quantization \citep{nagel2021white, guo2024decoupleq}.  
Consequently, the optimized model achieves performance comparable to the offline model described in \S\ref{subsec:clone} in both subjective and objective tests, with a significant reduction in latency, computation, and memory consumption, as shown in \autoref{tab:latency}.


\begin{table}[h]
\centering
\begin{tabular}{l|c|c|c|c|c}
\hline
\textbf{System} & \textbf{Latency ($\downarrow$)} & \textbf{RTF ($\downarrow$)} & \textbf{WER ($\downarrow$)} & \textbf{SIM ($\uparrow$)} & \textbf{CMOS ($\uparrow$)} \\\hline
Offline model & 1$\times$ & 1$\times$ & 1.518 & 0.763 & -\\
Deployed model & 0.028$\times$ & 0.132$\times$ & 1.518 & 0.763 & -0.02\\
\hline
\end{tabular}
\caption{\label{tab:latency}Comparison between the deployed model and the offline model.}
\end{table}

\section{Model extensions}

We further propose two extensions to the Seed-TTS model to enhance its performance and broaden its applicability. Initially, we introduce a self-distillation method designed to increase the controllability of timbre. Subsequently, we propose the use of reinforcement learning to holistically improve the model’s capabilities.

\subsection{Speech factorization by self-distillation}
\label{sec:vc}
Speech factorization refers to the process of decomposing speech into various independent, disentangled attributes. 
This feature allows TTS systems to flexibly synthesize speech with different combinations of timbre, prosody, and content from various speakers, which is crucial for applications like zero-shot voice conversion and factorized zero-shot TTS. Most prior approaches achieve attribute disentanglement through feature engineering \citep{chen2023streaming,wang2023lm,liu2021any,anastassiou2024voiceshop,lee2023hierspeech,choi2024dddm}, specific loss functions \citep{ju2024naturalspeech,jia2022zero}, or precise network architecture tuning \citep{qian2019autovc,popov2021diffusion,jia2022zero}. However, integrating these methods into a general-purpose speech generation system like Seed-TTS can be challenging.

We propose a self-distillation scheme to achieve attribute disentanglement. The core principle of this method is the creation of controlled speech pairs that share most information yet differ in one or a few specific target attributes. Utilizing such data pairs, along with minor updates to the model architecture, enables the Seed-TTS model to achieve high-quality attribute disentanglement. Given that Seed-TTS can produce high-quality zero-shot generation for nearly any speaker, generating these data pairs with varied target attributes is straightforward. In this report, we particularly highlight the process and results of timbre disentanglement.

We noticed that by introducing speaker perturbation into the diffusion module during Seed-TTS generation, we are able to obtain the synthetic speech with the same content and prosodic patterns but shifted timbres. We denote the original and timbre-altered sentences as $S_{ori}$ and $S_{alt}$, respectively. 

We retrain the diffusion model in the Seed-TTS system using these enhanced synthetic data pairs. Specifically, during training, the token extracted from $S_{alt}$ is used as the input of the network. A timbre reference extracted from $S_{ori}$ is also integrated as part of the diffusion input. 
The network is optimized to recover the vocoder embeddings extracted from $S_{ori}$. Notably, $S_{alt}$ and $S_{ori}$ share the same content and prosody but differ in timbre. To recover $S_{ori}$, the network must disregard the timbre embedded in the token sequence from $S_{alt}$ and rely solely on the provided timbre embedding. This approach allows us to modify the timbre using the additional timbre reference while preserving the original content and prosody. We find that this straightforward method enables the Seed-TTS system to achieve high-quality timbre disentanglement.

\begin{figure}[h]
\centering
\includegraphics[width=0.9\textwidth]{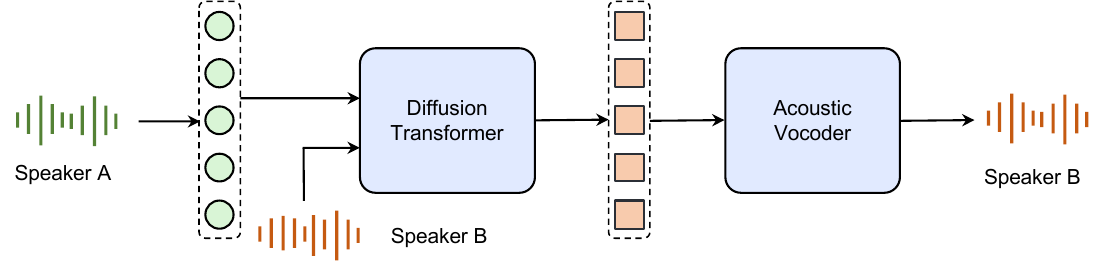}
\caption{\small The diagram for zero-shot voice conversion in Seed-TTS system.}
\label{fig:VCdiagram}
\end{figure}

We report the efficacy of the proposed disentanglement method through the zero-shot voice conversion (VC) task \citep{wang2023lm}. Zero-shot VC involves changing the speaker identity of source speech to a novel target timbre while preserving its spoken content. The diagram of the proposed VC pipeline is illustrated in \autoref{fig:VCdiagram}. In this setup, only the diffusion module of the Seed-TTS pipeline is involved in the VC experiments, as the content and prosody are dictated by the source speech.

We introduce a test set designed for zero-shot voice conversion evaluation based on the objective test set in \S\ref{subsec:clone}. Specifically, for each utterance, we randomly selected a non-matching speaker as the timbre reference. This test set configuration is released alongside the zero-shot ICL test set.
We conducted benchmarking experiments on this test set to assess the efficacy of our proposed method. We selected open-source SOTA methods for comparison, including HierSpeech++ \citep{lee2023hierspeech} and DiffVC \citep{popov2021diffusion}. Since these two methods only use English data for training, we restrict our evaluation to the English test subset.

The results are presented in \autoref{tab:vc}. We find our proposed self-distillation approach significantly improves the SIM metric through enhanced timbre disentanglement, while also being superior to pre-existing methods in all other dimensions. We have prepared a diverse range of audio examples, which can be found at \href{https://bytedancespeech.github.io/seedtts_tech_report/\#speech-factorization-samples}{this page}.

\begin{table}[h]
\centering
\begin{tabular}{l|c|c|c|c}
\hline
\multirow{2}{*}{\textbf{System}} & \multicolumn{2}{c|}{\textbf{Non-parallel ZH}} & \multicolumn{2}{c}{\textbf{Non-parallel EN}} \\ \cline{2-5} 
 & \textbf{WER ($\downarrow$)} & \textbf{SIM ($\uparrow$)} & \textbf{WER ($\downarrow$)} & \textbf{SIM ($\uparrow$)} \\ \hline
DiffVC \citep{popov2021diffusion}& - & - & 16.861 & 0.311 \\
HierSpeech++ \citep{lee2023hierspeech}& - & - & 5.469 & 0.387 \\ 
Seed-TTS (Ours) w/o self-distillation & 1.489 & 0.636 & 2.366 & 0.491 \\
Seed-TTS (Ours) with self-distillation & \textbf{1.216} & \textbf{0.791} & \textbf{2.121} & \textbf{0.753} \\ \hline
Before conversion & 1.254 & - & 2.143 & - \\ \hline
\end{tabular}
\caption{\label{tab:vc} Evaluation results on zero-shot voice conversion. The results of DiffVC \citep{popov2021diffusion} and HierSpeech++ \citep{lee2023hierspeech} are obtained via their respective released official checkpoints.}
\end{table}



\subsection{Preference biasing through reinforcement learning}
\label{sec:rl}
RL has proven to be an effective learning paradigm in text and image processing \citep{schulman2017proximal,rafailov2024direct,sutton1999policy,esser2024scaling,wallace2023diffusion}. Recent research has shown that Direct Preference Optimization (DPO) can be extended to music and speech generation \citep{cideron2024musicrl, zhang2024speechalign}. 

Inspired by these findings, we explore RL methods similar to those in previous studies \citep{ahmadian2024back,prabhavalkar2018minimum,wang2024transforming,sutton1999policy,schulman2017proximal} to enhance various aspects of Seed-TTS. We compare RL methods utilizing external reward models, such as Proximal Policy Optimization and REINFORCE, with those that do not, such as DPO. Our findings indicate that both approaches are effective. The former allows for clear control over specific speech attributes, while the latter benefits from a simpler implementation. In this report, we showcase the effectiveness of the former method.

Specifically, we use REINFORCE to fine-tune two versions based on the original zero-shot ICL model ($\text{Seed-TTS}_\text{ICL}$) using different reward functions: $\text{Seed-TTS}_\text{RL-SIM-WER}$, which uses the SIM and WER objective metrics as rewards to improve speaker similarity and robustness, and $\text{Seed-TTS}_\text{RL-SER}$, which uses the accuracy of the SER model as a reward to improve emotion controllability. 
We again use the same objective and subjective test sets mentioned in \S\ref{subsec:clone} to verify the contributions of RL in our system. Additionally, a new ``hard'' textual test set was prepared for the evaluation, consisting of 400 sentences with especially challenging patterns for autoregressive models such as word repetitions, tongue twisters, and so on. We report the results of objective and subjective evaluations in \autoref{tab:rl_icl_obj}, \autoref{tab:rl_icl_cmos}, and \autoref{tab:rl_emo_acc}. Audio examples can be found at \href{https://bytedancespeech.github.io/seedtts_tech_report/\#rl-samples}{this page}. 


\begin{table}[h]
\centering
\begin{tabular}{l|c|c|c}
\hline
\textbf{System} & \textbf{Test set} & \textbf{WER ($\downarrow$)} & \textbf{SIM ($\uparrow$)}  \\ \hline
\multirow{3}{*}{$\text{Seed-TTS}_\text{ICL}$} & ZH & 1.115 & 0.796 \\
 & EN & 2.249 & 0.762 \\ 
 & ``Hard'' & 7.585 & 0.776 \\ 
 \hline
\multirow{3}{*}{$\text{Seed-TTS}_\text{RL-SIM-WER}$} & ZH & \textbf{1.002} & \textbf{0.801} \\ 
 & EN & \textbf{1.945} & \textbf{0.766} \\ 
 & ``Hard'' & \textbf{6.423} & \textbf{0.782} \\
\hline 
\end{tabular}
\caption{\label{tab:rl_icl_obj} Objective evaluation results between $\text{Seed-TTS}_\text{RL-SIM-WER}$ and $\text{Seed-TTS}_\text{ICL}$.}
\end{table}

\begin{table}[h]
\centering
\begin{tabular}{l|c|ccc}
\hline
\multirow{2}{*}{\textbf{Systems}} & \multirow{2}{*}{\textbf{CMOS ($\uparrow$)}} & \multicolumn{3}{c}{\textbf{Preference (\%)}} \\ \cline{3-5} 
 &  & \textbf{Win} & \textbf{Tie} & \textbf{Loss} \\ \hline
$\text{Seed-TTS}_\text{RL-SIM-WER}$ vs. $\text{Seed-TTS}_\text{ICL}$ & \textbf{+0.14} & \textbf{44.1\%} & 25\% & 30.9\% \\ \hline
\end{tabular}
\caption{\label{tab:rl_icl_cmos} Subjective evaluation results between $\text{Seed-TTS}_\text{RL-SIM-WER}$ and $\text{Seed-TTS}_\text{ICL}$.}
\end{table}

\begin{table}[h]
\centering
\begin{tabular}{l|c|c|c|c}
\hline
\textbf{System} & \textbf{Angry} & \textbf{Happy} & \textbf{Sad} & \textbf{Surprise} \\ 
\hline
 $\text{Seed-TTS}_\text{ICL}$ & 0.46 & 0.44 & 0.53 & 0.13 \\
 $\text{Seed-TTS}_\text{RL-SER}$ & \textbf{0.91} & \textbf{0.8} & \textbf{0.78} & \textbf{0.82} \\
 \hline
\end{tabular}
\caption{\label{tab:rl_emo_acc} Comparison of the emotion control accuracy ($\uparrow$) between $\text{Seed-TTS}_\text{RL-SER}$ and $\text{Seed-TTS}_\text{ICL}$ in the zero-shot scenario using the emotion set from \autoref{sec:sft}.}
\end{table}


From \autoref{tab:rl_icl_obj} and \autoref{tab:rl_icl_cmos}, we observe the benefits of RL in both subjective and objective tests, resulting in improved stability and speaker similarity in the voice ICL task. In \autoref{tab:rl_emo_acc}, we find that although there is a decrease in emotion controllability in the zero-shot $\text{Seed-TTS}_\text{RL-SER}$ model compared to the speaker fine-tuned $\text{Seed-TTS}_\text{SFT}$ model in \S\ref{sec:sft}, the application of RL significantly improves the emotion controlled accuracy across various emotions compared to $\text{Seed-TTS}_\text{ICL}$. This enhancement highlights the efficacy of integrating RL techniques to boost performance in emotional expressiveness and control within speech synthesis models.


We observed reward hacking, which is a well-known issue for RL \citep{amodei2016concrete}, in our work. For example, in order to achieve a lower WER, the model tends to generate slower and more clearly pronounced utterances, which results in a sacrifice in naturalness. This observation aligns with the findings in \S\ref{subsec:clone}, where an excessively low WER often leads to more ``standardized'' but less natural speech.
Careful network tuning is required to achieve the optimal performance that balances these trade-offs afforded by RL. 

\subsection{Fully diffusion-based speech generation}

Language modeling and diffusion models are two main methodologies for multimedia generation. Several prior works directly compare their performance in image and video generation \citep{yu2023language}, yet we believe such a comparison for speech and audio generation remains limited. To further understand the characteristics of these two modeling approaches, we propose a variation of the Seed-TTS model based solely on diffusion, denoted as $\text{Seed-TTS}_\text{DiT}$. In this variation, we remove the dependency between the diffusion model and the acoustic tokenizer, such that the diffusion model directly converts Gaussian noise to the latent representation of the vocoder purely based on the input text.

We empirically find that including an additional duration prediction model as in \citet{jiang2023mega,ren2019fastspeech}, and \citet{le2024voicebox} results in reduced naturalness of synthesized speech. Therefore, in our modified design of $\text{Seed-TTS}_\text{DiT}$, we directly employ end-to-end processing within the diffusion model. As opposed to estimating phoneme-level durations, the model estimates the total duration of the generated speech beforehand. The model is then optimized to estimate the local alignment between audio and text. In this way, $\text{Seed-TTS}_\text{DiT}$ can dynamically adjust the duration of each phoneme, resulting in highly natural speech. 

We find $\text{Seed-TTS}_\text{DiT}$ is able to predict an appropriate total duration for input speech when trained properly. However, rather than training in this manner, we choose to directly provide the total duration to the model, which enables several additional desirable properties that may be used for content editing and speaking rate editing. To this end, during training the diffusion model receives the audio prompt, target text, and a clip of Gaussian noise with the total duration for each sample and predicts the latent representation of the generated speech with the same total duration, which is then transformed into a waveform by the vocoder. 

\begin{figure}[h]
\centering
\includegraphics[width=0.9\textwidth]{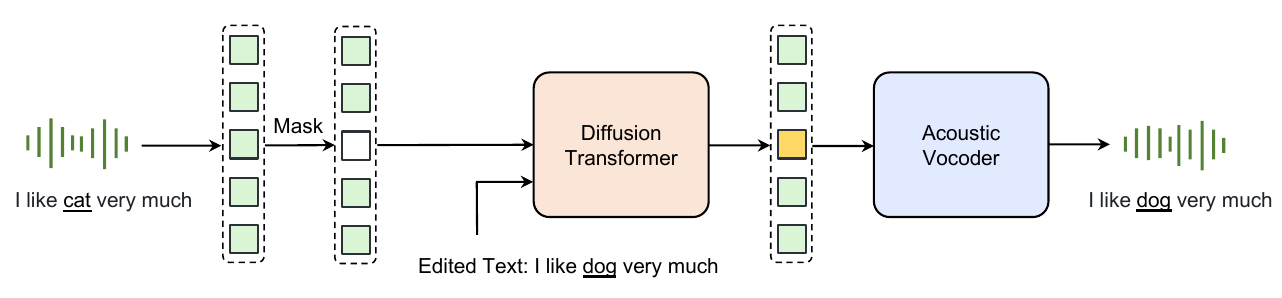}
\caption{\small Fully diffusion-based model $\text{Seed-TTS}_\text{DiT}$, supporting speech content editing. In this example, we replace the word ``cat'' in the original speech with the word ``dog''.}
\label{fig:contentedit}
\end{figure}

Compared with methods that employ next-token language modeling, the pure diffusion model enjoys a simpler pipeline. As a non-streaming model, $\text{Seed-TTS}_\text{DiT}$ naturally supports the application of content editing \citep{wang2023speechx,le2024voicebox,jiang2023mega}, as depicted in \autoref{fig:contentedit}. With that said, the language modeling approach has the advantage of streaming processing and the capability to integrate with the text-based language model.

We use the same test set as in \S\ref{subsec:clone} to evaluate $\text{Seed-TTS}_\text{DiT}$ on the zero-shot TTS task and present the evaluation results in \autoref{tab:dit}. We find that the fully diffusion-based method achieves superior performance in SIM while achieving similar results to $\text{Seed-TTS}_\text{ICL}$ in terms of WER. This finding indicates strong capability for sequence modeling inherent to the diffusion model.

\paragraph{Content editing and speaking rate editing.} We further evaluate $\text{Seed-TTS}_\text{DiT}$ on two speech editing tasks: content editing and speaking rate editing. We conduct these experiments using the ground truth counterpart of samples from the test set used in \S\ref{subsec:clone}. 

\begin{figure}[h]
\centering
\includegraphics[width=0.8\textwidth]{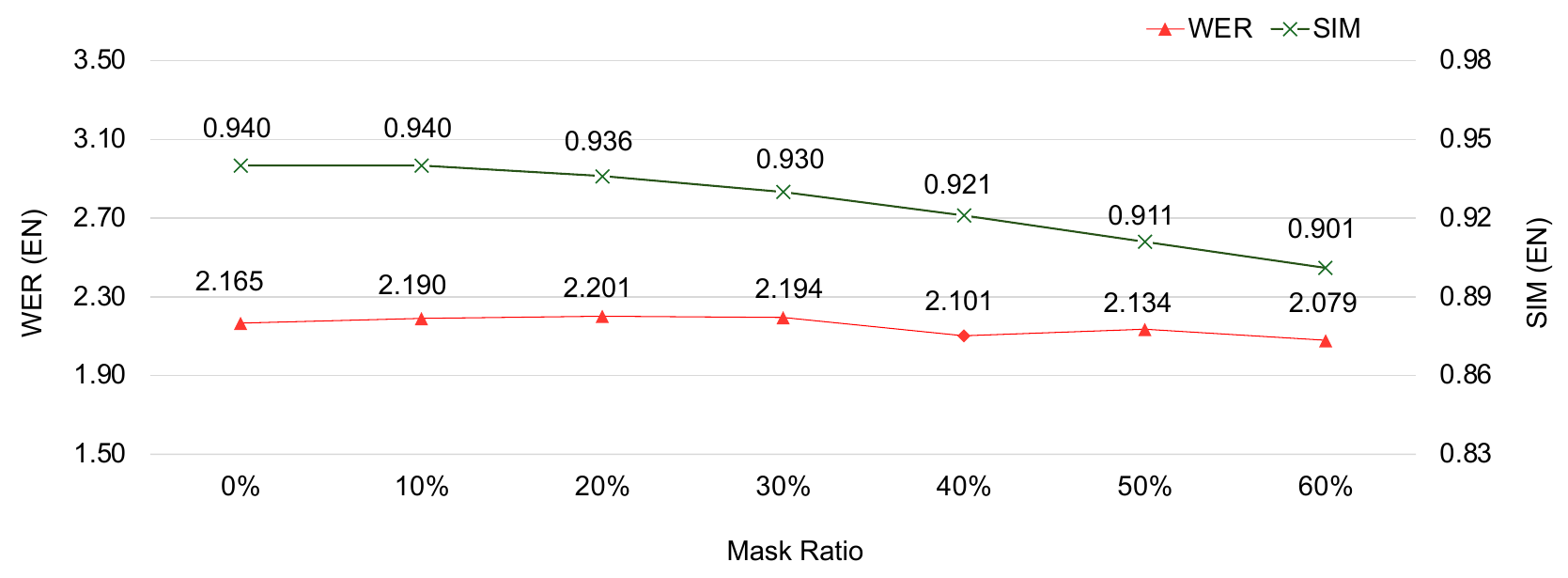}
\caption{\label{fig:edit_masked_rate} $\text{Seed-TTS}_\text{DiT}$ exhibits robustness across various masked rates in content editing.}
\end{figure}

In the content editing task, we mask a certain percentage of the audio and use the model to recover the masked portions based on the provided text for each test sample. We continue to employ WER and SIM as objective evaluation metrics. Specifically, we compute the SIM metric based on the recovered audio and the original audio to determine whether the recovered audio resembles the original speaker. The evaluation results are shown in \autoref{fig:edit_masked_rate}. We present diverse audio examples at \href{https://bytedancespeech.github.io/seedtts_tech_report/\#full-diffusion-samples}{this page}.

\begin{table}[h]
\centering
\begin{tabular}{l|c|c|c}
\hline
\textbf{System} & \textbf{Lang.} & \textbf{WER ($\downarrow$)} & \textbf{SIM ($\uparrow$)} \\\hline
Human & EN & 2.143& 0.730\\
Vocoder resynthesized & EN & 2.165&0.702\\
Seed-TTS$_\text{ICL}$ & EN&2.249&0.762\\
$\text{Seed-TTS}_\text{DiT}$ & EN&\textbf{1.733}&\textbf{0.790}\\
\hline
Human & ZH& 1.254& 0.750\\
Vocoder resynthesized & ZH & 1.342&0.733\\
Seed-TTS$_\text{ICL}$ &ZH &\textbf{1.115}&0.796\\
$\text{Seed-TTS}_\text{DiT}$ &ZH &1.178&\textbf{0.809}\\
\hline
\end{tabular}
\caption{\label{tab:dit}\small Objective evaluation results on zero-shot TTS. $\text{Seed-TTS}_\text{DiT}$ demonstrates superiority in both stability and speaker similarity.}
\end{table}

\begin{figure}[h]
\centering
\includegraphics[width=0.8\textwidth]{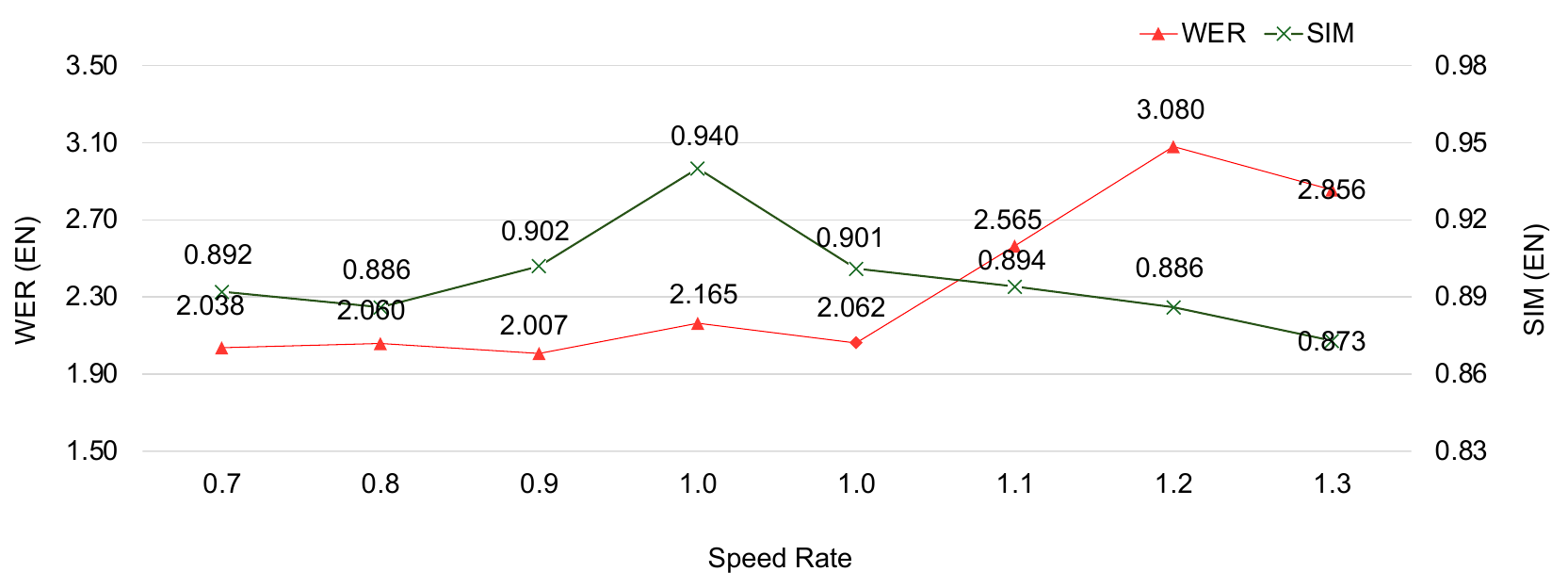}
\caption{\label{fig:edit_speed} $\text{Seed-TTS}_\text{DiT}$ is capable of synthesizing speech of different speeds with high speaker similarity. The WER shows a slight degradation when the speed rate is too high.}
\end{figure}

In the speaking rate editing task, we simply re-synthesize each test example with the modified total duration. Specifically, we obtain the final duration of the sentence by multiplying a speed rate with the original utterance duration. Identical to the content editing task, we utilize WER and SIM as objective evaluation metrics. The results are shown in \autoref{fig:edit_speed}. 

From our demonstration, it is evident that the model can automatically adjust the speaking rate solely based on different total durations. For example, when stretching speech into a longer total duration, the model will automatically insert silence at appropriate moments based on the input text or stretch the pronunciation of certain vowels while keeping the overall speaking rate within a natural range. In this way, the output speech produces improved naturalness and speaker similarity compared with traditional methods for these tasks that uniformly alter the speaking rate of the entire sentence.

\newpage

\section{Model applications, limitations, and safety}

The Seed-TTS model series, with its ability to create highly expressive and cross-lingual transferred speech, enables upgrades across several applications including voice chats, audio books, and content creation. Moreover, with its high-fidelity in-context learning, Seed-TTS enhances accessibility across language barriers and offers a potential solution to patients with speech impairments \citep{openai@voice}. As discussed in \S\ref{subsec:clone}, Seed-TTS also serves as a potential bridge to enhance and unify speech understanding and generation models. We demonstrate some potential applications at \href{https://bytedancespeech.github.io/seedtts_tech_report/\#applications-samples}{this page}.


Despite its capabilities, Seed-TTS has several limitations. Although emergent behavior is observed, the model sometimes has limitations in scenarios requiring nuanced emotion and contextual understanding. Additionally, despite being trained with a vast amount of data, there is still room for improvement in scenario coverage. For instance, the current Seed-TTS model does not perform well at singing or when given prompts containing background music or excessive noise, often generating inconsistent backgrounds, such as ignoring the music altogether.

Given the potential for harmful social impacts if misused, we implement multiple safety procedures in related products to prevent misuse throughout the development and potential deployment of this model. For example, we developed a multi-step verification method for spoken content and speaker timbre to ensure that enrollment audio contains only the voice of authorized users. Additionally, we implemented a multi-level watermarking scheme, which is mandatorily included at various levels in the created content, such as video background watermarks and watermarks in the content description. 



\newpage
\bibliographystyle{unsrtnat}
\bibliography{sample}

\newpage
\section{Authors (alphabetical order)}
\begin{multicols}{3}
\noindent
Philip Anastassiou \\
Jiawei Chen \\
Jitong Chen \\
Yuanzhe Chen\\
Zhuo Chen\\
Ziyi Chen\\
Jian Cong\\
Lelai Deng\\
Chuang Ding\\
Lu Gao\\
Mingqing Gong\\
Peisong Huang\\
Qingqing Huang\\
Zhiying Huang\\
Yuanyuan Huo\\
Dongya Jia\\
Chumin Li\\
Feiya Li\\
Hui Li\\
Jiaxin Li\\
Xiaoyang Li\\
Xingxing Li\\
Lin Liu\\
Shouda Liu\\
Sichao Liu\\
Xudong Liu\\
Yuchen Liu\\
Zhengxi Liu\\
Lu Lu\\
Junjie Pan\\
Xin Wang\\
Yuping Wang\\
Yuxuan Wang\\
Zhen Wei\\
Jian Wu\\
Chao Yao\\
Yifeng Yang\\
Yuanhao Yi\\
Junteng Zhang\\
Qidi Zhang\\
Shuo Zhang\\
Wenjie Zhang\\
Yang Zhang\\
Zilin Zhao\\
Dejian Zhong\\
Xiaobin Zhuang\\
\end{multicols}

\section{Acknowledgement}
We extend our deepest gratitude to the teams whose dedication and expertise were vital to the success of this project. Special thanks to our outstanding audio understanding team and engineering team for their technical prowess; our data teams, whose diligent efforts in data collection, annotation, and processing were indispensable; our project operation team for seamlessly providing guidance; and our evaluation team for their rigorous testing and insightful feedback. Each team’s unique contribution has been instrumental in bringing this research to fruition, and their collective efforts have been truly invaluable.

\end{document}